\documentclass[10pt,twocolumn,twoside]{IEEEtranTCOM}
\ifCLASSINFOpdf
\else
\fi
\usepackage{amsmath}
\newcommand{\subparagraph}{}
\usepackage{titlesec}
\usepackage{etoolbox}
\usepackage[linesnumbered,ruled,vlined]{algorithm2e}
\usepackage{enumerate}
\usepackage{graphicx}
\usepackage{enumitem}
\usepackage{amsmath}
\usepackage{algpseudocode}
\usepackage{amssymb}
\usepackage{times}
\usepackage{endnotes}
\usepackage{stfloats}
\usepackage{stkernel}
\usepackage{color}
\usepackage{enumerate}
\usepackage{array}
\usepackage{relsize}
\usepackage[belowskip=-10pt,aboveskip=3pt,footnotesize]{caption}
\usepackage{cite}
\renewcommand{\baselinestretch}{0.98}

\newcommand{\refeq}[1]{(\ref{#1})}
\newcommand{\tr}[1]{\textrm{#1}}
\newcommand{\jexp}[1]{e^{\jmath\left({#1}\right)}}
\newcommand{\nnsum}[2]{\underset{#1}{\overset{#2}{\sum}}}
\newcommand{\nsum}[2]{{\sum_{#1}^{#2}}}
\newcommand{\transp}{\mathrm{T}}
\newcommand{\nsigma}[2]{\sigma_{\mathrm{ #1}}^{{ #2} }}
\newcommand{\N}[1]{N_{\mathrm{ #1}}}

\newcommand{\T}[1]{T_{\mathrm{ #1}}}

\newcommand{\limit}[2]{\underset{{#1} \to {#2}} \lim}

\newcommand{\temp}[1]{\mathrm{I}_{\mathrm{#1}}}
\newcommand{\btemp}[1]{{\beta}_{\mathrm{#1}}}

\newcommand{\bw}[1]{\mathbf{w}_{#1}}
\newcommand{\w}[2]{w_{#1}^{(#2)}}
\newcommand{\tw}[2]{{w'}_{#1}^{(#2)}}
\newcommand{\tbw}[1]{{\mathbf{w'}}_{#1}}

\newcommand{\bThetahat}[1]{\hat{\mathbf{\Theta}}_{#1}}
\newcommand{\bTheta}[1]{\mathbf{\Theta}_{#1}}
\newcommand{\thetat}[2]{\theta_{#1}^{(#2)}}

\newcommand{\ghat}[1]{\hat{\mathbf{g}}_{#1}}
\newcommand{\bG}[1]{\mathbf{G}_{#1}}
\newcommand{\bg}[1]{\mathbf{g}_{#1}}
\newcommand{\g}[2]{g_{#1}^{(#2)}}

\newcommand{\del}{D}
\newcommand{\spow}{\sqrt{{P}}}
\newcommand{\pow}{{{P}}}
\newcommand{\by}[1]{\mathbf{y}_{#1}}
\newcommand{\y}[2]{y_{#1}^{(#2)}}

\newcommand{\bc}[1]{\mathbf{c}_{#1}}
\newcommand{\csym}[1]{c_{#1}}
\newcommand{\chat}[1]{\hat{{c}}_{#1}}

\newcommand{\nt}{\tilde{n}}
\newcommand{\phit}[2]{\phi_{#1}^{(#2)}}
\newcommand{\varphit}[1]{\varphi_{#1}}

\titlespacing*{\section}{0pt}{7pt plus 1pt minus 1pt}{7pt plus 1pt minus 1pt}

\begin{document}

\title{On the Impact of Oscillator Phase Noise on the Uplink Performance in a Massive MIMO-OFDM System}

\author{Rajet Krishnan, M. Reza Khanzadi, \IEEEmembership{Student~Member,~IEEE,} N. Krishnan,~\IEEEmembership{Member,~IEEE,} A. Graell i Amat, \IEEEmembership{Senior Member, IEEE,} T. Eriksson, N. Mazzali, \IEEEmembership{Member,~IEEE,} G. Colavolpe, \IEEEmembership{Senior Member, IEEE.}

\thanks{Rajet Krishnan, M. Reza Khanzadi, A. Graell i Amat, and T. Eriksson are with the Department
of Signals and Systems, Chalmers University of Technology, Gothenburg, Sweden (e-mail: \{rajet,
alexandre.graell, thomase\}@chalmers.se). N. Mazzali is with the Interdisciplinary Centre for Security, Reliability and Trust, University of Luxembourg (email: nicolo.mazzali@uni.lu). G. Colavolpe is with Dipartimento di Ingegneria dell' Informazione, University of Parma, Italy (e-mail: giulio.colavolpe@unipr.it). N. Krishnan is with Qualcomm, San Diego, California, USA (e-mail: nakrishn@qti.qualcomm.com).}%
\thanks{Research supported by the Swedish Research Council under grant \#2011-5961.}
}%
 \makeatletter
\patchcmd{\@maketitle}
  {\addvspace{0.5\baselineskip}\egroup}
  {\addvspace{-1\baselineskip}\egroup}
  {}
  {}
\makeatother

\markboth{IEEE Signal Processing Letters}%
{Under Review}

\maketitle% \thispagestyle{empty}
\vspace{-0.5cm}
\begin{abstract}

In this work we study the effect of oscillator phase noise on the uplink performance of a massive multiple-input multiple-output system. Specifically, we consider an orthogonal frequency-division multiplexing-based uplink transmission and analyze two scenarios: (a) all the base station (BS) antennas are fed by a common oscillator, and (b) each of the BS antennas is fed by a different oscillator. For the scenarios considered, we derive the instantaneous signal-to-noise ratio on each subcarrier and analyze the ergodic capacity when a linear receiver is used. Furthermore, we propose a phase noise tracking algorithm based on Kalman filtering that mitigates the effect of phase noise on the system performance.

\end{abstract}

\section{Introduction}
\label{sec:intro}

\PARstart{M}{assive}  multiple-input multiple-output (MIMO) is an emerging technology that will play a key role in the development of future energy-efficient, high-speed wireless networks  \cite{rusek13,marzetta10}. Systems that employ this technology, referred to as massive MIMO systems, involve the deployment of antenna-arrays that contain hundreds of antennas, which are expected to significantly increase the network throughput and also reduce power consumption \cite{larsson13}. However, the performance of these systems is severely limited by hardware impairments and the channel estimation accuracy \cite{emil13}.

In particular, massive MIMO systems that employ orthogonal frequency-division multiplexing (OFDM) are expected to be particularly sensitive to phase noise due to noisy local oscillators \cite{schenk05,ness06,fettweis07}. Phase noise in MIMO-OFDM systems results in two effects, namely, the common phase error (CPE)  and inter-carrier interference (ICI) \cite{marc95}. These impairments can severely affect the accuracy of the channel estimation and cause significant performance degradation. Importantly, phase noise causes the \textit{channel-aging phenomenon} \cite{heath13}, where the true channel, during the data transmission period, is significantly different from the channel estimate acquired during the training period. Hence, phase noise poses a serious challenge towards realizing the full potential promised by the massive MIMO theory. Prior work analyzing the effect of phase noise on a massive MIMO uplink transmission can be found in \cite{larsson12}, where single-carrier transmission is considered.

In this work, we analyze the impact of phase noise due to noisy local oscillators on the uplink performance for a massive MIMO-OFDM system that consists of a base station (BS) equipped with $M$ antennas, and a single-antenna user-equipment (UE). We consider two scenarios: (a) all the BS antennas are fed by a common oscillator (referred to as the CO scenario), and (b) each BS antenna is fed by a different oscillator (DO scenario). First, we propose a new MIMO-OFDM system model that accounts for both the CO and the DO scenarios, unlike the model in all prior works  \cite{schenk05,ness06,fettweis07}, which is valid only in the CO scenario. For both scenarios, we derive the instantaneous signal-to-noise (SNR) on each subcarrier for the number of BS antennas $M\rightarrow \infty$, and analyze the ergodic capacity when a maximum-ratio combining (MRC) receiver is used. Based on the instantaneous SNR derived, we study the effect of channel-aging in both scenarios. We then propose a phase noise tracking algorithm based on Kalman filtering that helps to mitigate the effects of phase noise due to CPE and channel-aging.

\section{System Model}
\label{sec:sys_mod}

Consider a single-user MIMO system with a BS and a UE, and assume that the UE is communicating with the BS in the uplink by means of an OFDM transmission, i.e, the information symbols are modulated over $\N{c}$ subcarriers with spacing $1/\T{s}$, where $\T{s}$ is the symbol duration (and hence the sampling period). During the uplink transmission, the UE transmits training- and data-OFDM symbols in non-overlapping time periods. Specifically, in the training period the UE transmits a pilot OFDM-symbol (where the information symbols present in all the subcarriers are known to the BS), which is used by the BS to estimate the channel links between its antennas and the UE, and to receive the transmitted data by means of an MRC receiver.

The channel between the BS antennas and the UE is assumed to be frequency-selective Rayleigh fading. The channel link for the $n$th subcarrier, $n \in {0,\ldots, \N{c}-1}$, between the UE and the $m$th BS antenna, $m \in {1,\ldots, M},$ is denoted by $\g{j,n}{m}$ in the $j$th time instant. Without loss of generality it is assumed that the large-scale fading component of the channel is unity, and $ \g{j,n}{m} \sim \mathcal{CN}(0,\nsigma{\mathrm{g}}{2})$, where $\nsigma{\mathrm{g}}{2}=1$. For both the CO and the DO scenarios, we consider that free-running oscillators are connected to the antennas \cite{colavolpe05}. In the $j$th time instant, the phase noise sample at the UE is denoted by $\varphit{j}$, and that at the $m$th BS antenna  by $\phit{j}{m}$,
\begin{IEEEeqnarray}{rCl}
\varphit{j}  &=& \varphit{j-1} + \Delta_{j}^{\varphi},\quad \Delta_{j}^{\varphi} \sim \mathcal{N}(0,\nsigma{\varphi}{2}),\\ \label{eq:wiener}
\phit{j}{m}  &=& \phit{j-1}{m} + \Delta_{j}^{\phi_m},\quad \Delta_{j}^{^{\phi_m}} \sim \mathcal{N}(0,\nsigma{\phi}{2}),
\end{IEEEeqnarray}
where  $\nsigma{\varphi}{2}$ and $\nsigma{\phi}{2}$ denote the phase noise increment variances at the UE and BS, respectively. Since the channel is constant within its coherence time, ${\varphit{j} + \phit{j}{m}}$ is the phase noise sample that impairs the link between the UE and the $m$th BS antenna. In the above discussion, it is implicitly assumed that Nyquist pulses are used for transmission, followed by matched filtering and sampling at every $\T{s}$ time period at the BS. Furthermore, in the sequel, when considering the CO scenario, we will drop the antenna index $m$ for notational convenience.

Let $i  \triangleq j = l\N{c}$, $l \in \mathbb{Z}^+$, then the frequency domain representation of the $l$th OFDM-symbol received at the BS antennas in the $i$th time instant is written as
%%\vspace{-1ex}
\begin{IEEEeqnarray}{rCl} \label{eq:sys_model}
\mathbf{y}_{i} = \spow \bTheta{i} \bG{i} \bc{i} + \mathbf{w}_{i}
\end{IEEEeqnarray}
where $P$ denotes the transmit power per subcarrier. Let  $P\triangleq 1/M^{\alpha}$, $\alpha > 0$, i.e., the UE transmit power is allowed to scale as $ 1/M^{\alpha}$, which is used to study the energy-spectral efficiency tradeoff in the system as in \cite{larsson12}. The received signal corresponding to all subcarriers is denoted by $\mathbf{y}_{i} = [\by{i,0}, \ldots, \by{i,\N{c}-1}]^{\transp}$, and $\by{i,n} = [\y{i,n}{1}, \ldots, \y{i,n}{M}]^{\transp}$ denotes the received signal at the $M$ receive antennas corresponding to the $n$th subcarrier. Furthermore,
\[ \bTheta{i} =  \left( \begin{array}{cccc}
\bTheta{i,0} & \bTheta{i,1} & \ldots & \bTheta{i,\N{c}-1} \\
\bTheta{i,-1} & \bTheta{i,0} & \ldots & \bTheta{i,\N{c} -2} \\
\vdots & \ddots & \ddots & \vdots \\
\bTheta{i,-\N{c}+1} & \bTheta{i,-\N{c}+2} & \ldots & \bTheta{i,0}
\end{array} \right)_{M\N{c}\times M\N{c}},\]
where $\bTheta{i,n}  = \mathrm{diag}\{ \thetat{i,n}{1}, \ldots,  \thetat{i,n}{M} \}$, and  $\thetat{i,n}{m} = \frac{1}{\N{c}} \nsum{\nt = 0}{\N{c}-1} \jexp{\frac{2\pi\nt n}{\N{c}}}\jexp{\varphit{i + \nt} + \phit{i + \nt}{m}}$, i.e., $\thetat{i,n}{m}$ is the $n$th Fourier coefficient of the $\N{c}$ discrete phase noise samples affecting an OFDM-symbol in each channel link. $\bG{i}  = \mathrm{diag}\{ \bg{i,0}, \ldots,  \bg{i,\N{c}-1} \}$ is the channel matrix of size $M\N{c}\times \N{c}$, and $\bg{i,n}  = [ \g{i,n}{1}, \ldots,  \g{i,n}{M} ]^{\transp}$. The symbols transmitted by the UE in each of the subcarriers are denoted by $\bc{i} = [\csym{i,0}, \ldots, \csym{i,\N{c}-1}]^{\transp}$, where $\csym{i,n} \in \mathcal{C}$, and $\mathcal{C}=\{-1,+1\}$ is considered to be  a binary phase-shift keying (BPSK) constellation set. The additive white Gaussian noise (AWGN) affecting the subcarriers is denoted by the vector $ \mathbf{w}_{i} = [\bw{i,0}, \ldots, \bw{i,\N{c}-1}]^{\transp}$ of dimension $M\N{c}\times 1$, where
$\bw{i,n} = [\w{i,n}{1}, \ldots, \w{i,n}{M}]^{\transp}$, and $\w{i,n}{m} \sim \mathcal{CN}(0,\nsigma{w}{2})$. It is worth noting that the system model in \refeq{eq:sys_model} accounts for both CO and DO scenarios in a MIMO-OFDM system, while that in prior work, e.g., \cite{schenk05,ness06,fettweis07}, is only applicable to the CO scenario.
%That is, $\thetat{i,n}{m}$ is the $n$th Fourier coefficient of the $\N{c}$ discrete phase noise samples affecting an OFDM-symbol in each channel link.
%for $m \in {1,\ldots, M}, n \in {0,\ldots, \N{c}-1}$,
\section{Instantaneous SNR and Ergodic Capacity}
\label{sec:analysis}

In this section, we analyze the instantaneous SNR that can be attained on each subcarrier when an MRC receiver is employed and $M\to \infty$. This SNR is used to evaluate the ergodic capacity \cite{marzetta10,diggavi02} in both the CO and the DO scenarios, and to study the effect of channel-aging. Without loss of generality, for the $0$th subcarrier, the received signal model at time instant $i=0$ is written as
\begin{IEEEeqnarray}{rCl} \label{eq:sys_model_sc}
\by{0,0} = \spow \underbrace{\bTheta{0,0}}_{\text{CPE}} \bg{0,0} \csym{0,0} + \underbrace{\spow \nsum{n = 1}{\N{c}-1} \bTheta{0,n} \bg{0,n} \csym{0,n}}_{\text{ICI}} + \bw{0,0},
\end{IEEEeqnarray}
where it can be seen that the phase noise manifests itself in the form of CPE that affects all the information symbols equally, and ICI.

Assume that the training period commences at time instant $i=0$, where the UE transmits one pilot OFDM-symbol. This is used by the BS to estimate the channel between its antennas and the UE for all subcarriers. Let the pilot symbols transmitted on all subcarriers in the OFDM-symbol be $+1 \in \mathcal{C}$. At the BS, the channel estimate $\ghat{0,0}$ is computed as $\ghat{0,0} = \by{0,0}$, which is also a scaled version of the maximum likelihood estimate of $\bTheta{0,0} \bg{0,0}$ under the assumption that ICI plus $\bw{0,0}$ in \refeq{eq:sys_model_sc} is approximately Gaussian distributed.

Now, consider that the data transmission period of the UE starts $\del$ time instants (i.e., $\del/\N{c}$ OFDM-symbol periods) after the training period, where $\del$ is lesser than the coherence time of the channel, implying that $\bg{D,0} = \bg{0,0}$. Note that in a multiuser MIMO system, this time delay typically arises because the UEs transmit pilot-OFDM symbols in orthogonal time slots during their training period, which is then followed by the data transmission period. Upon performing MRC on the signal received in the $0$th subcarrier in the $\del$th time instant, the transmitted information symbol is detected as
%\vspace{-1ex}
\begin{IEEEeqnarray}{rCl} \label{eq:mrc_sym}
\chat{\del,0} = \ghat{0,0}^*\by{\del,0} \triangleq \temp{SIG}  + \temp{ICI} + \temp{AWGN},
\end{IEEEeqnarray}
%%\vspace{-2ex}
% which is also the maximum likelihood estimate of the channel under the assumption that ICI plus $\tbw{i,0}$ is approximately Gaussian distributed
where $(\cdot)^{*}$ denotes the Hermitian transpose operator. In \refeq{eq:mrc_sym}, the desired signal component along with the effects of the CPE is denoted by $ \temp{SIG}$, the ICI component as $ \temp{ICI}$, and the AWGN component as $ \temp{AWGN}$. Their powers are computed in the sequel in order to derive the instantaneous SNR. From \refeq{eq:sys_model_sc} and \refeq{eq:mrc_sym}, the instantaneous power of the desired signal component $\temp{SIG} = \pow \bg{0,0}^*\bTheta{0,0}^* \bTheta{\del,0}\bg{\del,0}\csym{D,0} = P\nsum{m_1=1}{M}|\g{0,0}{m_1}|^2{\thetat{0,0}{m_1}}^*\thetat{D,0}{m_1}\csym{D,0}$ is evaluated as
%\vspace{-0ex}
%\begin{IEEEeqnarray}{rCl} \label{eq:i_sig}
%\temp{SIG} = \pow \bg{0,0}^*\bTheta{0,0}    \bTheta{\del,0}\bg{\del,0},
%\end{IEEEeqnarray}
%\vspace{-1ex}
\begin{IEEEeqnarray}{rCl} \label{eq:desired_pow}
&&\left|\temp{SIG}\right|^2 = \frac{1}{M^{2\alpha}} \left(\nsum{m_1=1}{M}|\g{0,0}{m_1}|^2{\thetat{0,0}{m_1}}^*\thetat{D,0}{m_1}\csym{D,0}\right)\nonumber\\&\cdot& \left(\nsum{m_2=1}{M}|\g{0,0}{m_2}|^2{\thetat{0,0}{m_2}}^*\thetat{D,0}{m_2}\csym{D,0}\right)^*\nonumber\\
%&=& \pow^2 \nnsum{m_1 = 1}{M}|\g{0,0}{m_1}|^4{\thetat{0,0}{m_1}}{\thetat{0,0}{m_1}}^*{\thetat{\del,0}{m_1}}{\thetat{\del,0}{m_1}}^* \vphantom{3^{3^{3}}}  \nonumber\\
%&&+ \pow^2 \nnsum{\stackrel{m_2=1,}{m_2 \neq m_1}}{M}|\g{0,0}{m_1}|^2|\g{0,0}{m_2}|^2{\thetat{0,0}{m_1}}^*{\thetat{0,0}{m_2}}{\thetat{\del,0}{m_1}}{\thetat{\del,0}{m_2}}^* \nonumber\\
 \label{eq:desired_pow_a}
&=&  \frac{1}{M^{2\alpha-1}} \nsum{m_1=1}{M}\frac{|\g{0,0}{m_1}|^4|{\thetat{0,0}{m_1}}|^2|{\thetat{\del,0}{m_1}}|^2}{M} \nonumber\\
& +&  \frac{M(M-1)}{M^{2\alpha}} \nnsum{m_1 = 1}{M}\nnsum{\stackrel{m_2=1,}{m_2 \neq m_1}}{M(M-1)} \nonumber\\&& \frac{|\g{0,0}{m_1}|^2|\g{0,0}{m_2}|^2{\thetat{0,0}{m_1}}^*{\thetat{0,0}{m_2}}{\thetat{\del,0}{m_1}}{\thetat{\del,0}{m_2}}^*}{M(M-1)}.
\end{IEEEeqnarray}
In \refeq{eq:desired_pow_a}, for large $M$, we have $M(M-1)\approx M^2$. From  the  law of large numbers (LLN), for any independent and identically distributed (i.i.d.) sequence of random variables (RVs) $\{a_1,\ldots,a_K\}, \{b_1,\ldots,b_K\}$ of length $K$, we have
\begin{IEEEeqnarray}{rCl} \label{eq:llnprodexp}
 \limit{K}{\infty} \nsum{k=1}{K}\frac{a_k b_k}{K} \rightarrow \mathbb{E}\{a b\} = \mathbb{E}\{a\}\mathbb{E}\{ b\},  \IEEEeqnarraynumspace
\end{IEEEeqnarray}
where $\mathbb{E}\{\cdot\}$ denotes the expectation operation. Applying \refeq{eq:llnprodexp} in \refeq{eq:desired_pow_a} results in
\begin{IEEEeqnarray}{rCl} \label{eq:desired_pow_b}
 \limit{M}{\infty} \left|\temp{SIG}\right|^2 &=& \limit{M}{\infty} \frac{2\temp{\mathrm{PN_1}}}{M^{2\alpha-1}}  +   \frac{ \temp{\mathrm{PN_2}}}{M^{2\alpha-2}},
\end{IEEEeqnarray}
since the fourth moment for a zero-mean complex Gaussian RV of unit variance can be computed as $\mathbb{E}\{|\g{0,0}{m_1}|^4\} = 2$ and $\mathbb{E}\{ |\g{0,0}{m_1}|^2|\g{0,0}{m_2}|^2\} = 1$.
%\begin{eqnarray} \label{eq:desired_pow_lim}% \IEEEeqnarraynumspace
%$\mathbb{E}\{|\g{0,0}{m_1}|^4\} = 8$ and $\mathbb{E}\{ |\g{0,0}{m_1}|^2|\g{0,0}{m_2}|^2\} = 1$.
%\end{eqnarray}
The term $\temp{\mathrm{PN_1}}$ in \refeq{eq:desired_pow_b} due to phase noise is evaluated for $n_1, n_2,n_3,n_4 \in \{0,\ldots,\N{c}-1\}$ as
%\begin{IEEEeqnarray}{rCl} \label{eq:desired_pow_pn1_a}
%\temp{\mathrm{PN_1}} &=& \frac{1}{{\N{c}}^4}\sum_{n_1} \sum_{n_2} \sum_{n_3} \sum_{n_4} \jexp{\varphit{\del + n_2} - \varphit{\del + n_4} }\jexp{\varphit{n_3} - \varphit{n_1} } \nonumber\\ && \cdot \jexp{\phi_{\del + n_2} - \phi_{\del + n_4} }\jexp{\phi_{n_3} - \phi_{ n_1} }, \text{for the CO scenario,} \\
%\label{eq:desired_pow_pn1_b}
%\temp{\mathrm{PN_1}} &=& \frac{1}{{\N{c}}^4}\sum_{n_1} \sum_{n_2} \sum_{n_3} \sum_{n_4}  \jexp{\varphit{\del + n_2} - \varphit{\del + n_4} }\jexp{\varphit{n_3} - \varphit{n_1} } \nonumber\\ && \cdot e^{\frac{-\nsigma{\phi}{2}}{2}(|n_4 - n_2|+|n_3-n_1|)}, \text{for the DO scenario},
%\end{IEEEeqnarray}
\begin{IEEEeqnarray}{rCl} \label{eq:desired_pow_pn1_a}
\temp{\mathrm{PN_1}} &=& \frac{1}{{\N{c}}^4}\sum_{n_1} \sum_{n_2} \sum_{n_3} \sum_{n_4} \jexp{\varphit{\del + n_2} - \varphit{\del + n_4} }\jexp{\varphit{n_3} - \varphit{n_1} } \nonumber\\ && \cdot \btemp{PN_1}(n_1,n_2,n_3,n_4)\\
\label{eq:desired_pow_pn1_b}
\btemp{PN_1} &=& \left\{ \begin{array}{ll}
         \jexp{\phi_{\del + n_2} - \phi_{\del + n_4} }\jexp{\phi_{n_3} - \phi_{ n_1} } & \mbox{CO scenario}\\
         e^{\frac{-\nsigma{\phi}{2}}{2}(|n_4 - n_2|+|n_3-n_1|)} & \mbox{DO scenario}\end{array} \right., \IEEEeqnarraynumspace
\end{IEEEeqnarray}
where \refeq{eq:desired_pow_pn1_b} is obtained by computing the characteristic function of the sum of the Gaussian RVs $\phit{n}{m} - \phit{n'}{m} = \nsum{\nt=n}{n'}\Delta_{\nt}^{\phi}$ from \refeq{eq:wiener}, for time indices $n, n'$. For the CO scenario, $\temp{\mathrm{PN_2}} = \temp{\mathrm{PN_1}}$, while for the DO scenario,
\begin{IEEEeqnarray}{rCl} \label{eq:desired_pow_pn2_b}
\temp{\mathrm{PN_2}} &=& \frac{1}{{\N{c}}^4}\sum_{n_1} \sum_{n_2} \sum_{n_3} \sum_{n_4} \jexp{\varphit{\del + n_2} - \varphit{\del + n_4} }\jexp{\varphit{n_3} - \varphit{n_1} } \nonumber\\ && \cdot e^{\frac{-\nsigma{\phi}{2}}{2}(|\del + n_2 - n_1|+|\del + n_3-n_4|)}.
\end{IEEEeqnarray}

The instantaneous power of the ICI term  in \refeq{eq:mrc_sym}, $\temp{ICI} \triangleq  \pow \nsum{n = 1}{\N{c}-1}  \bg{0,0}^*\bTheta{0,0}^* \bTheta{0,n} \bg{0,n} \csym{0,n}$, is derived as
%\vspace{-2ex}
\begin{IEEEeqnarray}{rCl} \label{eq:ici_pow}
& \limit{M}{\infty}&  \left|\temp{ICI}\right|^2 \nonumber \\ &=& \limit{M}{\infty}  \frac{1}{M^{2\alpha}}\left(\nsum{\nt_1= 1}{\N{c}-1}\nsum{m_1=1}{M} {\g{0,0}{m_1}}^*\g{D,\nt_1}{m_1}{\thetat{0,0}{m_1}}^*\thetat{D,\nt_1}{m_1}\csym{D,\nt_1}\right) \nonumber\\
&& \cdot \left(\nsum{\nt_2= 1}{\N{c}-1}\nsum{m_2=1}{M}{\g{0,0}{m_2}}^*\g{D,\nt_2}{m_2}{\thetat{0,0}{m_2}}^*\thetat{D,\nt_2}{m_2}\csym{D,\nt_2}\right)^* \\
\label{eq:ici_pow_b}
&=& \limit{M}{\infty}  \frac{M}{M^{2\alpha}} \nsum{\nt_1= 1}{\N{c}-1}\nsum{m_1=1}{M} \frac{|{\g{0,0}{m_1}}|^2|\g{D,\nt_1}{m_1}|^2|\thetat{0,0}{m_1}|^2|\thetat{D,\nt_1}{m_1}|^2}{M} \IEEEeqnarraynumspace\\
\label{eq:ici_pow_c}
&\overset{\triangle}{=}& \limit{M}{\infty}  \frac{1}{M^{2\alpha-1}}  \temp{PN_3},
\end{IEEEeqnarray}
%\vspace{-1ex}
where in \refeq{eq:ici_pow}, we use the fact that applying LLN  for $m_1\neq m_2$, or $\nt_1 \neq \nt_2$, results in
\begin{IEEEeqnarray}{rCl} \label{eq:ici_pow_lim}
&&\limit{M}{\infty} \nsum{\nt_2= 1}{\N{c}-1}\nsum{\nt_1= 1}{\N{c}-1}\nsum{m_1=1}{M} \nsum{m_2=1}{M} \frac{ {\g{0,0}{m_1}}^*{\g{0,0}{m_2}}\g{D,\nt_1}{m_1}{\g{D,\nt_2}{m_2}}^*}{M} = 0.\IEEEeqnarraynumspace
\end{IEEEeqnarray}
Furthermore, in \refeq{eq:ici_pow_c} the phase noise contribution in ICI, $\temp{PN_3}$, can be simplified as
%\vspace{-2ex}
\begin{IEEEeqnarray}{rCl} \label{eq:ici_pow_pn1_so}
\temp{\mathrm{PN_3}} &=&  \frac{1}{{\N{c}}^4}\sum_{n_1} \sum_{n_2} \sum_{n_3} \sum_{n_4}  \jexp{\varphit{\del + n_2} - \varphit{\del + n_4} }\jexp{\varphit{n_3} - \varphit{n_1} } \nonumber\\ && \cdot \btemp{PN_3}(n_1,n_2,n_3,n_4)\nnsum{\nt_1=1}{\N{c}-1} \jexp{\frac{2\pi (n_2 - n_4) \nt_1}{\N{c}}}, \\
%\temp{\mathrm{PN_3}} &=&  \frac{1}{{\N{c}}^4}\sum_{n_1} \sum_{n_2} \sum_{n_3} \sum_{n_4}  \jexp{\varphit{\del + n_2} - \varphit{\del + n_4} }\jexp{\varphit{n_3} - \varphit{n_1} } \nonumber\\ && \cdot\nnsum{\nt_1=1}{\N{c}-1} \jexp{\frac{2\pi (n_2 - n_4) \nt_1}{\N{c}}}\\
%&=& \frac{1}{{\N{c}}^2}\nnsum{n_1=0}{\N{c}-1} \nnsum{n_3=0}{\N{c}-1} \jexp{\varphit{n_3} - \varphit{n_1} } \jexp{\phi_{n_3} - \phi_{ n_1} }
 \btemp{PN_3} &=& \left\{ \begin{array}{ll}
          \jexp{\phi_{\del + n_2} - \phi_{\del + n_4} }\jexp{\phi_{n_3} - \phi_{ n_1} } & \mbox{CO scenario},\\
         e^{\frac{-\nsigma{\phi}{2}}{2}(|n_4 - n_2|+|n_3-n_1|)}& \mbox{DO scenario}.\end{array} \right.
\end{IEEEeqnarray}
%\vspace{-2ex}
%In the DO scenario,
%\begin{IEEEeqnarray}{rCl} \label{eq:ici_pow_pn1_mo}
%\temp{\mathrm{PN_3}} &=&  \frac{1}{{\N{c}}^4}\sum_{n_1} \sum_{n_2} \sum_{n_3} \sum_{n_4}  \jexp{\varphit{\del + n_2} - \varphit{\del + n_4} }\jexp{\varphit{n_3} - \varphit{n_1} } \nonumber\\ && \cdot e^{\frac{-\nsigma{\phi}{2}}{2}(|n_4 - n_2|+|n_3-n_1|)}\nnsum{\nt_1=1}{\N{c}-1} \jexp{\frac{2\pi (n_2 - n_4) \nt_1}{\N{c}}} %\\
%%&=& \frac{1}{{\N{c}}^2}\nnsum{n_1=0}{\N{c}-1} \nnsum{n_3=0}{\N{c}-1} \jexp{\varphit{n_3} - \varphit{n_1} } e^{\frac{\nsigma{\phi}{2}}{2}(|n_3-n_1|)}
%\end{IEEEeqnarray}

The instantaneous AWGN power in \refeq{eq:mrc_sym} is given as
\begin{IEEEeqnarray}{rCl} \label{eq:awgn_pow}
& \limit{M}{\infty} & \left|\temp{AWGN}\right|^2  = \limit{M}{\infty} \frac{1}{M^{\alpha}} \left| \bg{0,0}^*{\bTheta{0,0}}^*\bw{D,0}  \right. \nonumber\\  &&\left.  + M^{\frac{\alpha}{2}}\tbw{0,0}^*\bw{D,0} + \nnsum{\nt_1 = 0}{\N{c}-1}\tbw{0,0}^* \bTheta{D,\nt_1}\bg{D,\nt_1} \csym{D,\nt_1} \right|^2 \\
\label{eq:awgn_pow_b}
&=& \limit{M}{\infty} \frac{1}{M^{\alpha}}  \left| \nnsum{m_1 = 1}{M} {\g{0,0}{m_1}}^*{\thetat{0,0}{m_1}}^*{\w{D,0}{m_1}} \right. \nonumber\\  && \left. + M^{\frac{\alpha}{2}} \nnsum{m_1 = 1}{M} {\tw{0,0}{m_1}}^* \w{D,0}{m_1} +  \nnsum{m_1 = 1}{M} \nnsum{\nt_1 = 0}{\N{c}-1} {\tw{0,0}{m_1}}^* \thetat{D,\nt_1}{m_1} \g{D,\nt_1}{m_1}\right|^2 \nonumber\\
\label{eq:awgn_pow_c}
&=& \limit{M}{\infty}\frac{1}{M^{\alpha}} \nnsum{m_1 = 1}{M} |{\g{0,0}{m_1}}|^2 |{\thetat{0,0}{m_1}}|^2 |{\w{D,0}{m_1}}|^2 \nonumber\\ &&+ \limit{M}{\infty}M \nnsum{m_1 = 1}{M} \frac{|{\tw{0,0}{m_1}}|^2 |\w{D,0}{m_1}|^2}{M} \nonumber\\ &&+ \limit{M}{\infty}\frac{1}{M^{\alpha}}  \nnsum{\nt_1 = 0}{\N{c}-1} \nnsum{m_1 = 1}{M} | {\tw{0,0}{m_1}}|^2 |\thetat{D,\nt_1}{m_1}|^2 |\g{D,\nt_1}{m_1}|^2 \\
%&=& \limit{M}{\infty}\frac{1}{M^{\alpha}-1} \nnsum{m_1 = 1}{M} |{\g{0,0}{m_1}}|^2 |{\thetat{0,0}{m_1}}|^2 |{\w{D,0}{m_1}}|^2 \nonumber\\ &+& \limit{M}{\infty}M \nnsum{m_1 = 1}{M} \frac{|{\tw{0,0}{m_1}}|^2 |\w{D,0}{m_1}|^2}{M} \nonumber\\ &+& \limit{M}{\infty}\frac{1}{M^{\alpha}}  \nnsum{\nt_1 = 0}{\N{c}-1} \nnsum{m_1 = 1}{M} | {\tw{0,0}{m_1}}|^2 |\thetat{D,\nt_1}{m_1}|^2 |\g{D,\nt_1}{m_1}|^2 \\
\label{eq:awgn_pow_d}
&=& \limit{M}{\infty}\frac{1}{M^{\alpha-1}} \temp{PN_4} \nsigma{\mathrm{w}}{2} + \limit{M}{\infty}M  (\nsigma{w}{2} + \frac{1}{M^{\alpha}} \nsigma{\textrm{ICI}}{2})\nsigma{{w}}{2} \nonumber\\ &&+ \limit{M}{\infty}\frac{1}{M^{\alpha-1}}   (\nsigma{w}{2} + \frac{1}{M^{\alpha}} \nsigma{\textrm{ICI}}{2}) \temp{PN_5}.
\end{IEEEeqnarray}
In \refeq{eq:awgn_pow} $\tbw{0,0} \triangleq \spow \nsum{n = 1}{\N{c}-1} \bTheta{0,n} \bg{0,n} \csym{0,n} + \bw{0,0}$, and ${w'}_{0,0}^{(m)}\sim \mathcal{CN}(0,\nsigma{w}{2} + \frac{1}{M^{\alpha}} \nsigma{\textrm{ICI}}{2})$, where the ICI term  in \refeq{eq:sys_model_sc} is approximated as a Gaussian RV, which is independent of $\w{0,0}{m}$ and its variance  $\frac{1}{M^{\alpha}}  \nsigma{\textrm{ICI}}{2}$ is computed as in \cite{fettweis07}. The cross terms in \refeq{eq:awgn_pow} are uncorrelated, and applying LLN to them leads to the result in \refeq{eq:awgn_pow_b}. Finally, by applying LLN to \refeq{eq:awgn_pow_b}, the result in \refeq{eq:awgn_pow_d} is obtained. The term $\temp{PN_4}$ in \refeq{eq:awgn_pow_d} is derived as
%\vspace{-0ex}
%\begin{IEEEeqnarray}{rCl} \label{eq:awgn_pow_pn1_so}
%\temp{\mathrm{PN_4}} &=&  \frac{1}{{\N{c}}^4}\sum_{n_1} \sum_{n_2} \jexp{\varphit{n_2} - \varphit{n_1} } \jexp{\phi_{n_2} - \phi_{n_1} }, \text{ CO scenario}\IEEEeqnarraynumspace\\
%\temp{\mathrm{PN_4}} &=& \frac{1}{{\N{c}}^4}\sum_{n_1} \sum_{n_2} \jexp{\varphit{n_2} - \varphit{n_1} }e^{\frac{-\nsigma{\phi}{2}}{2}|n_2 - n_1|},  \text{DO scenario}. \IEEEeqnarraynumspace
%\end{IEEEeqnarray}
\begin{IEEEeqnarray}{rCl} \label{eq:awgn_pow_pn4_so}
\temp{\mathrm{PN_4}} &=& \frac{1}{{\N{c}}^4}\sum_{n_1} \sum_{n_2} \sum_{n_3} \sum_{n_4} \jexp{\varphit{\del + n_2} - \varphit{\del + n_4} }\jexp{\varphit{n_3} - \varphit{n_1} } \nonumber\\ && \cdot \btemp{PN_4}(n_1,n_2,n_3,n_4)\\
\label{eq:desired_pow_pn4_b}
 \btemp{PN_4} &=& \left\{ \begin{array}{ll}
        \jexp{\varphit{n_2} - \varphit{n_1} } \jexp{\phi_{n_2} - \phi_{n_1} } & \mbox{CO scenario}\\
         \jexp{\varphit{n_2} - \varphit{n_1} }e^{\frac{-\nsigma{\phi}{2}}{2}|n_2 - n_1|} & \mbox{DO scenario}\end{array} \right.. \IEEEeqnarraynumspace
\end{IEEEeqnarray}
Likewise, the contribution due to phase noise $\temp{PN_5}$ is derived as
%\vspace{-2ex}
\begin{IEEEeqnarray}{rCl} \label{eq:awgn_pow_pn2_so}
&& \temp{\mathrm{PN_5}} =  \frac{1}{{\N{c}}^2} \sum_{n_1} \sum_{n_2} \jexp{\varphit{\del + n_2} - \varphit{\del + n_1} } \jexp{\phi_{\del + n_2} - \phi_{\del + n_1} } \nonumber\\  && \cdot \sum_{\nt_1}  \jexp{\frac{2\pi (n_2 - n_1) \nt_1}{\N{c}}} =  \nnsum{n_1=0}{\N{c}-1} \frac{1}{{\N{c}}} = 1,  \text{for the CO scenario}. \IEEEeqnarraynumspace\\
\label{eq:awgn_pow_pn2_do}
&&\temp{\mathrm{PN_5}} = \frac{1}{{\N{c}}^2} \sum_{n_1} \sum_{n_2}  \jexp{\varphit{\del + n_2} - \varphit{\del + n_1} }e^{\frac{-\nsigma{\phi}{2}}{2}(|n_2 - n_1|)}    \nonumber\\  &&\cdot \sum_{\nt_1} \jexp{\frac{2\pi (n_2 - n_1) \nt_1}{\N{c}}} =  \nnsum{n_1=0}{\N{c}-1} \frac{1}{{\N{c}}} = 1, \text{for the DO scenario}.\IEEEeqnarraynumspace
\end{IEEEeqnarray}
For deriving the results in \refeq{eq:awgn_pow_pn2_so} and \refeq{eq:awgn_pow_pn2_do}, the orthogonality of complex exponentials is used as in \cite{fettweis07}.

Using the results from \refeq{eq:desired_pow_b}, \refeq{eq:ici_pow_c} and \refeq{eq:awgn_pow_d}, we analyze the instantaneous SNR  \cite{ness04 ,syrjala13}, written as the ratio of the power of the desired signal and the signal power due to AWGN and ICI, as $M\rightarrow \infty$,
%\begin{IEEEeqnarray}{rCl} \label{eq:snr_final}
%&& \limit{M}{\infty} \mathsf{SNR} \nonumber\\ && =
%\limit{M}{\infty} \frac{\frac{2\temp{\mathrm{PN_1}}}{M^{2\alpha-1}}  + \frac{ \temp{\mathrm{PN_2}}}{M^{2\alpha-2}}}{\frac{\temp{PN_3}}{M^{2\alpha-1}}  + \frac{\temp{PN_4} \nsigma{\mathrm{w}}{2}}{M^{\alpha-1}}  +  M\nsigma{w}{4} +  \frac{\nsigma{\textrm{ICI}}{2}\nsigma{\mathrm{w}}{2}}{M^{\alpha-1}}  +  \frac{\nsigma{w}{2} \temp{PN_5}}{M^{\alpha-1}}  + \frac{\nsigma{\textrm{ICI}}{2} \temp{PN_5}}{M^{2\alpha-1}} }\IEEEeqnarraynumspace\\
%\label{eq:snr_final_b}
%&&= \frac{ \temp{\mathrm{PN_2}} }{ \nsigma{\mathrm{w}}{4}},
%\end{IEEEeqnarray}
\begin{IEEEeqnarray}{rCl}\label{eq:snr_final}
&& \limit{M}{\infty} \mathsf{SNR} \nonumber\\ &=&
\limit{M}{\infty} \frac{\frac{2\temp{\mathrm{PN_1}}}{M^{2\alpha-1}}  + \frac{ \temp{\mathrm{PN_2}}}{M^{2\alpha-2}}}{\frac{\temp{PN_3}}{M^{2\alpha-1}}  + \frac{\temp{PN_4} \nsigma{\mathrm{w}}{2}}{M^{\alpha-1}}  +  M\nsigma{w}{4} +  \frac{\nsigma{\textrm{ICI}}{2}\nsigma{\mathrm{w}}{2}}{M^{\alpha-1}}  +  \frac{\nsigma{w}{2} \temp{PN_5}}{M^{\alpha-1}}  + \frac{\nsigma{\textrm{ICI}}{2} \temp{PN_5}}{M^{2\alpha-1}} }\nonumber\\
\\
\label{eq:snr_final_b}
% &=& \left\{ \begin{array}{lll}
%        0 & \alpha > 1/2\\
%                         \frac{ \temp{\mathrm{PN_2}} }{ \nsigma{\mathrm{w}}{4}} & \alpha = 1/2 \\
%        \infty & \alpha < 1/2 \end{array} \right.. \IEEEeqnarraynumspace
 &=& \left\{
        0:  \alpha > 1/2; \frac{ \temp{\mathrm{PN_2}} }{ \nsigma{\mathrm{w}}{4}}: \alpha = 1/2; \infty: \alpha < 1/2  \right\}. \IEEEeqnarraynumspace
\end{IEEEeqnarray}
In \refeq{eq:snr_final_b}, it can be seen that decreasing $\alpha$ (below $1/2$) increases the spectral efficiency, but reduces the energy efficiency of the system. Finally, the ergodic capacity per subcarrier for the system in \refeq{eq:mrc_sym} is evaluated as
\begin{IEEEeqnarray}{rCl} \label{eq:erg_cap}
C_{\mathrm{erg}} &=& \mathbb{E} \{\log_{2}\left(1 +  \mathsf{SNR} \right)\}.
\end{IEEEeqnarray}

The following deductions can be made about $\mathsf{SNR}$ in \refeq{eq:snr_final} and \refeq{eq:snr_final_b}.
\begin{itemize}[leftmargin=*]
\item For both the CO and the DO scenarios, an array gain of $\mathcal{O}(\sqrt{M})$ is achievable for an uplink transmission impaired by phase noise. Furthermore, by letting $\alpha < 1/2$, the AWGN noise power is reduced to zero \refeq{eq:snr_final_b}, leaving behind only noise due to the CPE (from the local oscillator at the BS and the UE) and also pilot contamination (if present, as in typical multiuser  massive MIMO systems \cite{marzetta10}).
\item The ICI term due to phase noise is reduced to zero \refeq{eq:snr_final_b} when $M$ is large and $\alpha \leq 1/2$. Hence, phase noise compensation techniques need to be designed only to suppress the CPE component, which is present in $\temp{\mathrm{PN_2}}$.
\item For both the CO and DO scenarios, the phase noise from the UE is not averaged out for large $M$ \refeq{eq:desired_pow_pn1_b}, \refeq{eq:desired_pow_pn2_b}. However, for the DO scenario, the phase noise from the different oscillators at the BS is averaged to a deterministic value that depends on $\nsigma{\phi}{2}$. This averaging effect does not happen in the CO scenario for the phase noise at the BS.
\item The channel estimate computed in the training period becomes irrelevant in the data transmission period due to the random phase drift in the elapsed time $\del$ (channel-aging). However, the time elapsed $\del$ only affects the desired signal power in the DO scenario and does not affect in the CO scenario \refeq{eq:desired_pow_pn1_b}, \refeq{eq:desired_pow_pn2_b}. This is in line with the observation that, following a linear receiver, the received signal in the DO scenario experiences amplitude distortion due to phase noise \cite{rajet12}.
\end{itemize}

%\begin{IEEEeqnarray}{rCl} \label{eq:exp_desired_pow_pn2}
%\mathbb{E} {\temp{\mathrm{PN_2}}} &=& \frac{1}{{\N{c}}^4}\nnsum{n_1=0}{\N{c}-1} \nnsum{n_2=0}{\N{c}-1} \nnsum{n_3=0}{\N{c}} \nnsum{n_4=0}{\N{c}-1} e^{\frac{-\nsigma{\varphi}{2}}{2}(|n_2 - n_4|+|n_3-n_1|)} \nonumber\\ &\cdot&  e^{\frac{-\nsigma{\phi}{2}}{2}(|n_2 - n_4|+|n_3-n_1|)} \text{ for CO,}\\
%\mathbb{E} {\temp{\mathrm{PN_2}}} &=& \frac{1}{{\N{c}}^4}\nnsum{n_1=0}{\N{c}-1} \nnsum{n_2=0}{\N{c}-1} \nnsum{n_3=0}{\N{c}-1} \nnsum{n_4=0}{\N{c}-1} e^{\frac{-\nsigma{\varphi}{2}}{2}(|n_2 - n_4|+|n_3-n_1|)} \nonumber\\ &\cdot&  e^{\frac{-\nsigma{\phi}{2}}{2}(|\del + n_2 - n_1|+|\del + n_3-n_4|)} \text{ for DO}.
%\end{IEEEeqnarray}
\section{MIMO-OFDM Phase Noise Compensation and MRC}
\begin{figure}[!t]
\begin{center}
\includegraphics[width = 3.5in, keepaspectratio=true]{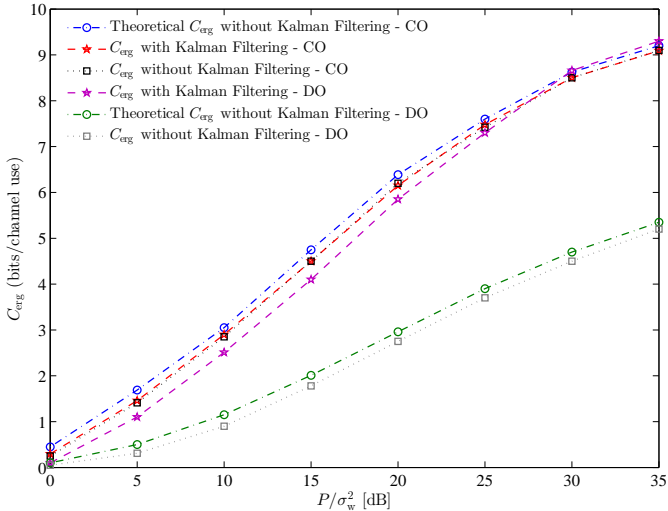}
\caption{Ergodic Capacity $C_{\tr{erg}}$ for the common oscillator (CO) and the different oscillators (DO) scenarios, where $\N{c}=64$, $M=100,$ $\nsigma{\varphi}{} = \nsigma{\phi}{}= 2^{\circ}$, and $D= 1280$. }
\label{fig:fig_mami1}
\end{center}
\end{figure}

As seen in the previous section, the random phase drift due to the CPE \refeq{eq:sys_model_sc} during the elapsed time $\del$ causes channel-aging, wherein $\ghat{0,0}$ becomes irrelevant in the data transmission period. In this section, a Kalman filter is designed to continuously track the CPE, which is then used to compensate for the channel-aging effect.

In order to continuously track the CPE, the training period of the UE is extended by $\del$ time instants (i.e., $\del/\N{c}$ OFDM-symbol periods). In a multiuser MIMO system, this corresponds to the scenario where all the UEs train simultaneously by using code-based orthogonal pilot sequences (like Walsh-Hadamard codes \cite{minn07}). Furthermore, in the training period the transmit power per subcarrier is rescaled by a factor of $\N{c}/D$, resulting in the following system model in the $i$th time instant (for $i>0$ and lesser than the channel coherence time),
\begin{IEEEeqnarray}{rCl} \label{eq:sys_model_filter}
\by{i,0} &=& \sqrt{\frac{P\N{c}}{\del}} \bTheta{i,0} \bg{i,0} \csym{i,0}  + \tbw{i,0}.
\end{IEEEeqnarray}
This is then followed by the data transmission period, where the transmit power is scaled back to $\pow$.

The CPE is approximated as a first-order auto-regressive (AR) process  \cite{colavolpe05},
\begin{IEEEeqnarray}{rCl}
%&&\jexp{\varphit{i + \N{c} + \nt} + \phit{i + \N{c} + \nt}{m}} =  \jexp{\varphit{i + \nt} + \phit{i +  \nt}{m}} e^{\jmath \nnsum{\nt_1 = 0}{\N{c}-1} (\Delta_{i + \nt + \nt_1}^{\varphi} + \Delta_{i + \nt + \nt_1}^{\phi_m})} \nonumber\\
%&& \nnsum{\nt = 0}{\N{c}-1} \jexp{\varphit{i + \N{c} + \nt} + \phit{i + \N{c} + \nt}{m}}  \nonumber\\ && = \nnsum{\nt = 0}{\N{c}-1} \jexp{\varphit{i + \nt} + \phit{i +  \nt}{m}} e^{\jmath \nnsum{\nt_1 = 0}{\N{c}-1} (\Delta_{i + \nt + \nt_1}^{\varphi} + \Delta_{i + \nt + \nt_1}^{\phi_m})} \\
\label{eq:filter_rkn_b}
&& \thetat{i+\N{c},0}{m} \approx \rho \thetat{i,0}{m} + {{v}^{(m)}_{i,0}}, \rho = e^{-\frac{\nsigma{\phi}{2} + \nsigma{\varphi}{2} }{2}},
\end{IEEEeqnarray}
where  ${v}^{(m)}_{i,0}\sim \mathcal{CN}(0,\N{c}(1- \rho^2))$, and $\rho$ is determined such that the variance of ${{v}^{(m)}_{i,0}}$ is minimized. Thus, the  state model is written as
\begin{IEEEeqnarray}{rCl}
\label{eq:sys_model_filter_b}
\bTheta{i+\N{c},0} &=& \rho \bTheta{i,0} + {\mathbf{v}_{i,0}}.
\end{IEEEeqnarray} % The problem of jointly estimating the CPE and the channel has been addressed extensively in prior works \cite{ness06,fettweis07}.

In order to apply the standard Kalman filter equations to the observation model in \refeq{eq:sys_model_filter} and the state model in \refeq{eq:sys_model_filter_b}, the knowledge of  $\bg{i,0}$ is required, which is not perfectly known at the receiver. Here, we assume perfect knowledge of $\bg{i,0}$ as in \cite{ness04,dhahir10}. The Kalman filter produces an estimate of the CPE, denoted by ${\bThetahat{i,0}}$. The estimate ${\bThetahat{i,0}}$ is combined with $\ghat{0,n},$ $n \in \{0,\ldots,\N{c}-1\}$, to produce the compensated channel estimate for any subcarrier, i.e., $\ghat{i,n} = \bThetahat{i,0}\ghat{0,n}$, which is then used for performing MRC reception.
%, the estimate $\ghat{0,0}$ is used as its true value
%In order to apply the standard Kalman filter equations to \refeq{eq:sys_model_filter}, the knowledge of $\bg{i,0}$ is required. The channel $\bg{i,0}$ is not perfectly known at the receiver, and the problem of jointly estimating the CPE and the channel has been addressed extensively in prior works \cite{ness06,fettweis07}. However, in this work, we assume perfect knowledge of $\bg{i,0}$ as in \cite{ness04,dhahir10}, and we focus exclusively on tracking the CPE in order to compensate for the channel-aging effect. The knowledge of $\csym{i,0}$ is also required in \refeq{eq:filter_rkn}, which is facilitated by inserting pilot symbols at  fixed subcarrier positions \cite{fettweis07}. Without loss of generality, let the pilot symbols be inserted in the $0$th subcarrier. The Kalman filter produces an estimate of the CPE, denoted by ${\bThetahat{i,0}}$. The estimate ${\bThetahat{i,0}}$ is combined with $\ghat{0,n},$ $n \in \{0,\ldots,\N{c}-1\}$, to produce the compensated channel estimate for any subcarrier, i.e., $\ghat{i,n} = \bThetahat{i,0}\ghat{0,n}$, which is then used for performing MRC reception.

We remark that the CPE can also be tracked in the data transmission period based on pilot symbols that are inserted at fixed subcarrier positions \cite{fettweis07}. However, in a multiuser system, pilot-based phase noise tracking methods will be affected by pilot contamination if the pilot symbols transmitted by the different users interfere with each other. This interference can be alleviated by appropriately designing pilot sequences \cite{minn07, marzetta13}. %The impact of pilot contamination on the system performance during the training and the data transmission periods is mitigated by appropriately designing pilot sequences for different users \cite{marzetta13}, which is beyond the scope of this work.

\section{Simulation Results and Discussion}
\label{sec:sim_res}

We simulate the system model described in Sec. \ref{sec:sys_mod} and numerically evaluate the ergodic capacity in order to analyze the uplink performance. Furthermore, we study the effectiveness of the designed Kalman filter to compensate for the channel-aging effect. The number of subcarriers per OFDM-symbol is set to $\N{c}= 64$, and the number of BS antennas is fixed to $M=100$. The phase noise increment standard deviations are set to $\nsigma{\varphi}{} = \nsigma{\phi}{}= 2^{\circ}$, which corresponds to a strong phase noise scenario. The time elapsed between the training and the data transmission periods is set as $D/\N{c} =20$ OFDM-symbol periods, i.e., $\del = 1280$. The results are shown in Fig.~\ref{fig:fig_mami1}, from which the following observations are made.
\begin{itemize}[leftmargin=*]
\item The ergodic capacity $C_{\mathrm{erg}}$  \refeq{eq:erg_cap} evaluated based on the SNR derived in \refeq{eq:snr_final} is observed to match with that obtained in the simulations, implying that the averaging effect due to receive diversity happens for even small values of $M$.
\item In the DO scenario, when the proposed channel-aging compensation is not applied, there is a significant degradation in the performance. However, in the CO scenario, both the uncompensated and the compensated systems exhibit similar performances.
\item Applying channel-aging compensation, the instantaneous SNR achieved in the CO scenario is higher than that in the DO scenario at low SNR, despite the averaging effect of the phase noise in the latter. This is because of the relatively higher estimation error incurred in tracking $M$ phase noise processes at the BS in the DO scenario, while in the CO scenario there are $M$ observations for a single phase noise process \cite{emil13}. However, the instantaneous SNR in the DO scenario becomes better than that in the CO scenario as the estimation error decreases with decrease in $\nsigma{w}{2}$, as seen at around $\frac{P}{\nsigma{w}{2}}=25$ dB in Fig.~\ref{fig:fig_mami1}.
%\item The impact of phase noise on the uplink performance is not significantly affected by the number of subcarriers  $\N{c}$. This is unlike in traditional OFDM transmissions, where an increase in $\N{c}$ results in the increased ICI, which degrades performance.
\end{itemize}

%\section{Conclusions}
%\label{sec:concl}

%\section*{APPENDIX A}  % use *-form to suppress numbering
%\section*{Derivation of the SPA Messages and Computation of their Parameters}
%\label{sec:app_a}

\bibliographystyle{IEEEbib}

\end{document}